\documentclass{aa}
\usepackage{graphicx}
\usepackage{natbib}
\usepackage{txfonts}
\bibpunct{(}{)}{;}{a}{}{,}

\def\I{\rm {\scriptsize I}}
\begin{document}
\title{Discovery of the eclipse in the symbiotic binary Z Andromedae}

\author{A. Skopal}

\institute{Astronomical Institute, Slovak Academy of Sciences,
           059\,60 Tatransk\'a Lomnica, Slovakia\\
           \email{skopal@ta3.sk}
          }

\date{Received; accepted}

\abstract{
Our photometric observations of the symbiotic binary Z\,And during 
its recent (2000 -- 2003) active phase revealed a minimum in the 
$U,~B$ and $V$ light curves (LC) at the position of the inferior 
conjunction of its cool component (the orbital phase $\varphi = 0$). 
This fact and the behaviour of colour indices suggest that 
the minimum was due to the eclipse of the active hot object by 
the red giant. 
Physically plausible fit of the eclipse profile and a precise 
analysis of the spectral energy distribution (SED) in the ultraviolet 
continuum suggest a disk-like structure for the hot object during 
active phases. The present knowledge of fundamental parameters 
of the system limits the orbital inclination $i$ to 
$76^{\circ} - 90^{\circ}$. 
The presence of the Rayleigh attenuated far-UV continuum at 
$\varphi \sim 0$ during quiescent phase confirms the very high 
inclination of the Z\,And orbit. 
         }

\maketitle
%
%

\section{Introduction}

Z\,And is the prototype of the symbiotic stars. Its brightness 
variation has been recorded from 1887 
\citep[e.g.][and reference therein]{matt}. 
The historical LC shows several active phases during which 
fluctuations ranges in amplitude from a few tenths of a magnitude 
to about 3 magnitudes \citep[e.g.][]{fl94}. 
During active phases the hot component expands in radius and 
becomes significantly cooler. A mass-outflow from the active star
is indicated directly by a broadening of emission lines and/or 
by profiles of the P-Cygni type 
\citep[e.g. Fern\'andez-Castro et al. (1995), 
hereafter FC95;][]{ss41}. 
The quiescent phase of Z\,And is characterized by a complex 
wave-like brightness variation as a function of the orbital phase 
\citep[e.g.][]{sk01a}. 
Based on all available primary minima in the LC from 1900 to 1996, 
\citet{sk98} determined their ephemeris as 
\begin{equation}
JD_{\rm Min} = 2\,414\,625.2 + 757.5(\pm 0.5)\times E,
\end{equation}
which is identical (within uncertainties) with timing of the 
inferior spectroscopic conjunction of the giant star in the system 
(Mikolajewska \& Kenyon 1996, hereafter MK96). 
We use this ephemeris further in this paper. 

The binary of Z\,And composes of a late-type, M4.5\,\I\I\I, giant 
\citep{ms99} 
with 
$M_{\rm g} \sim 2\,{\rm M_{\odot}}$,
$R_{\rm g} \sim 85 - 140\,{\rm R_{\odot}}$ 
and 
$L_{\rm g} \sim 880\,{\rm L_{\odot}}$ 
\citep[Nussbaumer \& Vogel 1989, hereafter NV89;][]{mk96,ttt03}. 
The hot component is probably a magnetic accreting white dwarf 
\citep{sb99} 
with 
$M_{\rm h} \sim 0.5 - 1\,{\rm M_{\odot}}$,
$L_{\rm h} \approx 900 - 2\,500\,{\rm L_{\odot}}$,
and 
$T_{\rm eff} \sim 10^{5}$\,K, 
surrounded by an ionized nebula \citep{mk96,f-c95}. 

Z\,And is considered to be a non-eclipsing binary. 
\citet{mk96} suggested the orbital inclination 
$i \approx 50^{\circ} - 70^{\circ}$ as a compromise between 
the behaviour of the He\,\I\I\,1640 emission -- showing a deep minimum 
around the giant's inferior conjunction -- and the lack of eclipses 
in the LC. Also \citet{ss97}, based on repeated polarimetric 
measurements, determined the orbital inclination to 
47$^{\circ}\pm 12^{\circ}$. 

In this {\em Letter} we report the first detection of the eclipse 
in the LC, which provides direct evidence for a very high 
inclination of the orbital plane of Z\,And. 

\section{Observations}

Our photometric $U,~B,~V,~R$ measurements of Z\,And were performed
in the standard Johnson system using single-channel photoelectric
photometers mounted in the Cassegrain foci of 0.6-m reflectors at
the Skalnat\'{e} Pleso and Star\'{a} Lesn\'{a} observatories. 
Z\,And was measured with respect to the comparison star 
SAO\,53150 ($V$ = 8.99, $B-V$ = 0.41, $U-B$ = 0.14, $V-R$ = 0.16) 
which was checked by the previous standard star SAO\,35642 
($V$ = 5.30, $B-V = -0.06$, $U-B = -0.15$, $V-R = -0.04$). 
To cover gaps in photoelectric observations we also used visual 
magnitude estimates gathered by the members of Association 
Francaise des Observateurs d'Etoiles Variables (AFOEV), which 
are available from the CDS. 
In addition, we compared a few V-CCD observations, which are 
available from the VSNET database. 
Figure~\ref{fig_1} shows the LCs covering the recent 
active phase of Z\,And. A table containing the data till 2001 
December and other details about photoelectric observations 
can be found in \citet{sk_al02}. 

Finally, we also used the low resolution spectra available from 
the final IUE archive (see Sect. 3.3). They were taken on 
06/04/84 (SWP22684 + LWP03099), 
24/12/85 (SWP27370 + LWP07370), 
11/07/86 (SWP28655 + LWP08585) and 
03/02/88 (SWP32845 + LWP12614). 
We dereddened the spectra with $E_{\rm B-V}$ = 0.30 \citep{nv89} by 
using the extinction curve of \citet{ccm89}. 

\section{Results}

\subsection{Evidence of the eclipse}

\begin{figure}[t]
   \resizebox{\hsize}{!}{\includegraphics[angle=-90]{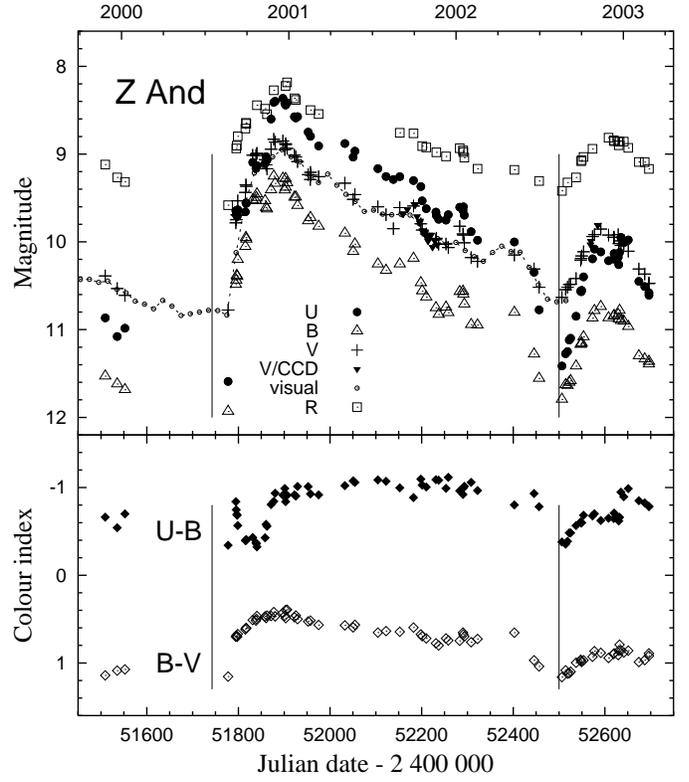}}
   \caption{
The $UBVR$ photometry of Z\,And covering the current 
active phase. The vertical lines denote the position of the inferior 
spectroscopic conjunction of the giant according to the ephemeris 
given by Eq. (1). Small circles connected with a broken line 
represent 20-day means of the visual estimates from CDS. 
}
\label{fig_1}
\end{figure}

Figure~\ref{fig_1} shows the $UBVR$ LCs of Z\,And covering its 
current outburst, which started in 2000 September \citep{sk_al00}. 
The star's brightness reached a maximum at the beginning 
of 2000 December, after which it has been gradually decreasing. 
The latest observations from the end of 2002 showed a transient 
small increase of the activity. 
The most pronounced feature in the LC is a deep minimum, which 
occurred around $JD~2\,452\,500$. We summarize its basic 
observational characteristics as follows: 

1. Position of the minimum agrees perfectly with the time of 
the inferior spectroscopic conjunction of the cool component 
in the binary (cf. Eq. 1). 

2. At the mid of the minimum ($JD~2\,452\,500$), Z\,And became 
considerably redder than before/after its beginning/end: The 
colour indices were identical with those observed prior to the 
outburst (cf. bottom panel of Fig.~\ref{fig_1}) and its brightness 
was very close to these values. 

These facts imply that the minimum was due to the eclipse 
of the active hot star by the red giant. 
%
%
%
%
\begin{figure}
   \resizebox{\hsize}{!}{\includegraphics[angle=-90]{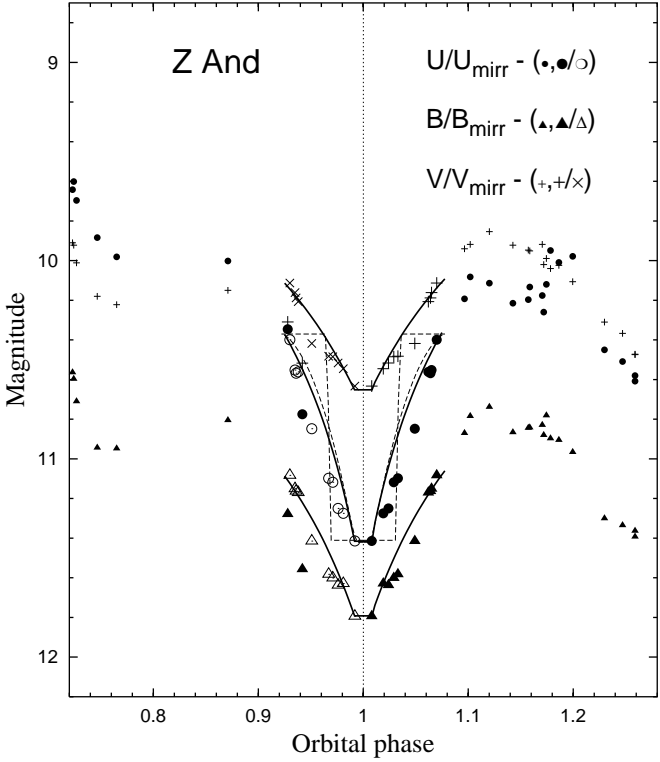}}
   \caption{
Comparison of the observed and the modeled eclipse profile. Solid 
thick line corresponds to the model of an extended belt, which is 
subject to the eclipse. This model suggests a disk-like structure 
of the hot active object and limits the orbital inclination to 
76$^{\circ}$ -- 90$^{\circ}$. Models with a spherical shape of 
the hot component (dashed lines) do not match observations 
(the rectangular shape) or are not physically plausible 
(the broad dashed profile). See Sect. 3.2 for more details. 
}
\label{fig_2}
\end{figure}

\subsection{Geometry of the eclipsed object}

The observed eclipse profile is basically of a V-type. However, 
its shape cannot be accurately determined, because of (i) a poor 
coverage of the descending branch by observations, (ii) an influence 
of a strong nebular contribution, which makes the top parts of 
the profile broader, and (iii) an enhanced activity during 
the time when the hot object was arising up from the eclipse. 
Second point is given by a strong nebular contribution in 
the spectrum during active phases (Sect. 3.3), and the third point 
is indicated by a higher level of the star's brightness 
after the eclipse then prior to it, although a reverse behaviour 
was expected due to a general decrease from the major outburst. 
The basic V-type profile of the eclipse and its position are also 
supported by continuous visual observations 
from CDS (Fig.~\ref{fig_1}). 

Therefore, to compare better the observed profile to a model, 
we mirrored the data from the ascending branch of the eclipse 
with respect to the time of the spectroscopic conjunction at 
the epoch $E$ = 50 (JD~2\,452\,500.2, Eq. 1). Figure~\ref{fig_2} 
shows the result, which suggests that radii of both 
the eclipsing and the eclipsed object are relatively large and 
comparable each to other. Based on fundamental parameters 
published by NV89, MK96, and \citet{ttt03}, we derived 
the range of the giant's radius as 
$R_{\rm g} \sim (0.17 - 0.32)\,A$ 
and limited the size of the eclipsed object by its Roche lobe 
radius, $R_{\rm ecl}^{\rm L}$, as 
$R_{\rm ecl} < R_{\rm ecl}^{\rm L} \sim (0.26 - 0.32)\,A$ 
($A$ stands for the separation between the centers of mass 
of the binary components). 
First we tried to match the eclipse profile for a spherical 
shape of the hot object by using a code of \citet{kv94}. 
However, its luminosity and temperature during active phases, 
$L_{\rm h} \approx 2\,500 - 3\,600\,L_{\sun}$ 
and 
$T_{\rm h} = 5 - 3\,10^{4}$\,K \citep{f-c95,ttt03},
imply an effective radius only of 
$R_{\rm ecl} \approx 2\,R_{\sun}$, 
which cannot reproduce the observed eclipse profile 
(Fig.~\ref{fig_2}). Contrary, a large sphere 
($R_{\rm ecl} \approx 100\,R_{\sun}$), which can reproduce formally 
the eclipse profile, has an unrealistically high luminosity 
($L_{\rm h} \approx 10^{5}\,L_{\sun}$). Therefore, we rejected 
models with the spherical shape of the eclipsed object. 
Second, we matched the observed profile by a simple model of 
a uniformly radiating belt around the central star as the 
eclipsed object by using own code. Such the model can simulate, 
for example, the geometry of an accretion disk seen edge-on, 
which is extended at its outer parts. Here we obtained 
the following solutions: 

(i) If the linear size of the giant's stellar disk that eclipses 
the belt, $R_{\rm g}^{\rm E}$, is larger than $R_{\rm ecl}$, the 
resulting models are determined by $R_{\rm ecl}/A$ = 0.21, 
$R_{\rm g}/A$ = 0.26\,($i$\,=\,90$^{\circ}$) -- 
0.32\,($i$\,=\,79$^{\circ}$) 
and the belt radiates 63\%, 49\% and 40\% of the total light in 
the $U$, $B$ and $V$ band, respectively. 

(ii) For $R_{\rm g}^{\rm E} < R_{\rm ecl}$, the models correspond to 
$R_{\rm ecl}/A$ = 0.26, 
$R_{\rm g}/A$ = 0.21\,($i$\,=\,90$^{\circ}$) -- 
0.32\,($i$\,=\,76$^{\circ}$) 
and the belt contributes 78\%, 61\% and 50\% of the total light 
in the $U$, $B$ and $V$ band, respectively.
Figure~\ref{fig_2} shows example given by parameters 
$R_{\rm g}/A = R_{\rm ecl}/A$ = 0.26 and $i = 81^{\circ}$
of the case (ii). 

Analysis of this section suggests that the hot eclipsed object 
during the activity has a disk-like structure. 
This result is supported independently by our analysis of 
the SED, which we present in the following section. 
%
%
\begin{figure}
   \resizebox{\hsize}{!}{\includegraphics[angle=-90]{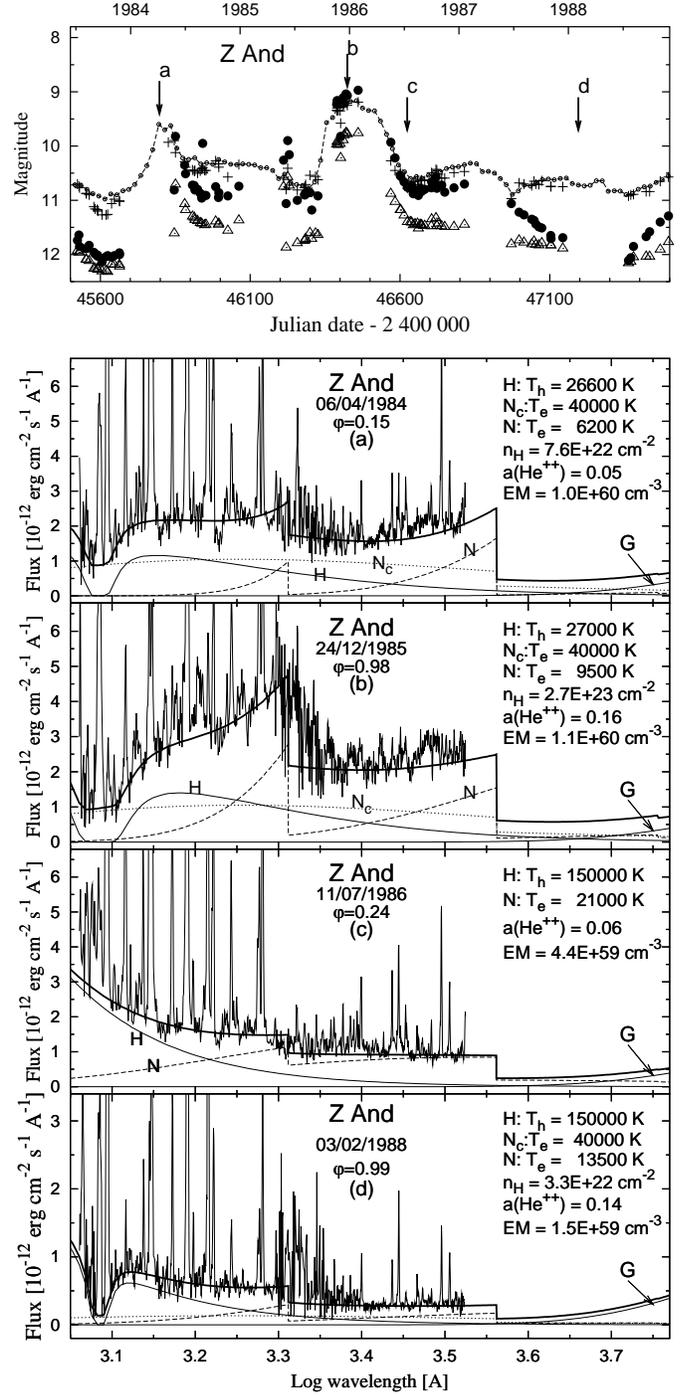}}
   \caption{
Reconstructed SED in the UV continuum of Z\,And during its active 
and quiescent phases (panels $\sf a, b$ and $\sf c, d$, respectively).
Solid thin lines (H, G) represent stellar components of radiation 
from the hot object and the red giant ($T_{\rm g}$ = 2\,800\,K). 
Dashed and dotted lines (N, N$_{\rm c}$) represent the nebular 
components of radiation and the solid thick line is the resulting 
modeled continuum (Sect. 3.3). Corresponding parameters are 
displayed at the top-right corner. Emission measure was calculated 
for $d$ = 1.12\,kpc. Denotation of LCs in the top panel 
is the same as in Fig.~\ref{fig_1}.
Double component temperature spectrum, which develops during 
the activity (panels {\sf a} and {\sf b}), suggests a disk-like 
structure of the hot object, which is seen approximately edge-on. 
The Rayleigh attenuated far-UV continuum, seen even during 
quiescence at $\varphi \sim 0$, confirms the high inclination 
of~the~orbital~plane~(panel~{\sf d}). 
}
\label{fig_3}
\end{figure}

\subsection{SED in the ultraviolet continuum}

The effect of eclipses of a multi-component source of radiation 
depends on a physical displacement and radiative contributions 
of its individual components in the system. In the case that 
a significant fraction of radiation at the wavelength under 
consideration comes from the region, which is subject to eclipse, 
a minimum in the LC is well observable. In the contrary case, 
the eclipse effect is very faint. In addition, the displacement 
and relative contributions of emission regions in a symbiotic 
system are a strong function of the level of the activity. 
Therefore, to demonstrate this situation for Z\,And, we use 
our method of a precise modeling of its ultraviolet continuum 
during both the active and the quiescent phase. 

%
%
In our procedure we assume that the observed flux, 
$F_{\lambda}$ [erg\,cm$^{-2}$\,s$^{-1}$\,$\AA^{-1}$], is given 
by superposition of fluxes from a hot stellar object and a nebula. 
We approximate the former by blackbody radiation, which can be 
Rayleigh attenuated, and for the latter we assume the f-b plus 
f-f transitions in fully ionized hydrogen and 
\mbox{helium. So, we express the observed flux as}
\begin{equation}
F_{\lambda}\, =\, k_{\rm h}\times \pi B_{\lambda}(T_{\rm h})
                  e^{-n_{\rm H}\sigma_{\lambda}^{\rm R}}\, +\, 
              k_{\rm N}\times\varepsilon_{\lambda}(T_{\rm e}, a). 
\end{equation}
The first term represents radiation of the blackbody at 
a temperature $T_{\rm h}$, which is attenuated by neutral 
atoms of hydrogen on the line of sight due to the Rayleigh 
scattering process ($n_{\rm H}$ [cm$^{-2}$] is the column 
density of H atoms, $\sigma_{\lambda}^{\rm R}$ [cm$^{2}$] 
is the Rayleigh scattering cross-section for atomic hydrogen 
and $k_{\rm h}$ is a dimensionless scaling factor). The second 
term represents the nebular radiation given by the volume 
emission coefficient, $\varepsilon_{\lambda}$ 
[erg\,cm$^{3}$\,s$^{-1}$\,$\AA^{-1}$], for fully ionized 
hydrogen and helium including all acts of recombination (f-b 
transitions) and thermal bremsstrahlung (f-f transitions). 
The coefficient is scaled to the observed nebular flux with 
the factor $k_{\rm N} = EM/4\pi d^{2}$ [cm$^{-5}$] ($EM$ is 
the emission measure and $d$ is the distance to the object) 
and was calculated by the same way as used by \citet{sk01b}. 
Finally, $a = n({\rm He^{++}})/n({\rm H})$ is the abundance of 
doubly ionized helium and $T_{\rm e}$ is electron temperature. 
%
%

First, we determined fluxes in the continuum at selected wavelengths. 
Because of noise, numerous emission lines and absorption 
features we preferred an estimate by eye. Second, we 
calculated a grid of models given by Eq. (2) for 6 parameters, 
$k_{\rm h},~T_{\rm h},~n_{\rm H},~k_{\rm N},~T_{\rm e},~a$, 
and selected that giving the best fit to the observed fluxes 
under conditions of the least square method. Figure~\ref{fig_3} 
shows a comparison of the observed spectra and the resulting 
models during active and quiescent phases. 
%
%
The resulting UV continuum from {\em active phases} 
(panels {\sf a}, {\sf b} of Fig.~\ref{fig_3}) consists of 
two components of radiation of a very different temperature 
regimes: A relatively cool stellar component 
($T_{\rm h} \sim 27\,000$\,K), which is Rayleigh attenuated at
the far-UV, and a very hot radiation, which gives rise to highly 
ionized emission lines and the nebular continuum. The latter 
consists of two -- low plus high temperature -- components. 
The hot nebula had to be added to fit the far-UV continuum. 
With analogy to AR\,Pav, the electron temperature of 40\,000\,K 
was adopted here \citep[cf.][]{sk03}. 
%
%
In the {\em quiescent phase} (panels {\sf c}, {\sf d} of 
Fig.~\ref{fig_3}), the stellar component of radiation corresponds 
to very high temperature $T_{\rm h} \sim $150\,000\,K \citep{nv89}, 
the low temperature nebular emission dominates the near-UV 
region and the hot coronal emission is very faint. Around 
the inferior conjunction of the giant, the far-UV continuum 
is Rayleigh attenuated. 
%
%
Below we summarize results of this analysis, which confirm 
those obtained by modeling the eclipse profile: 

(i) 
Double component temperature spectrum with the Rayleigh attenuated 
stellar component of radiation, which develops during 
{\em active phases}, suggests a disk-like structure for 
the eclipsed object seen approximately edge-on. Such configuration 
implies that the radiation from the hot central star 
($T_{\rm h} \sim 150\,000$\,K) is absorbed and diffused in 
the disk in the direction to the observer, and thus transformed to 
the observed $T_{\rm h} \sim 27\,000$\,K, but it is free in 
directions to the poles giving rise to the nebular emission there 
(panels~~{\sf a},~{\sf b}~of~Fig.~\ref{fig_3}). 

(ii)
The presence of the Rayleigh attenuated far-UV continuum at 
$\varphi \sim 0$ even during the {\em quiescent} phase is a 
consequence of the high orbital inclination -- a typical feature 
for eclipsing symbiotic binaries during quiescent phases 
\citep{inv89}. In this case the hot stellar radiation is scattered 
on the neutral atoms of hydrogen in the giant's wind at positions 
around $\varphi \sim 0$, but~it~is~not~present~at~$\varphi \ga
0.1$~(panels~~{\sf c},~{\sf d}~of~Fig.~\ref{fig_3}).

(iii)
At the maximum and during quiescence the nebular emission 
dominates the optical, which means that a significant fraction 
of the total hot object radiation can be physically displaced 
from the central hot star. 
By other words, a fraction of the radiation removed from the line 
of sight due to the eclipse can be very small, which precludes 
arising the eclipse effect in the LC. The eclipse effect can be 
observed only at specific brightness phases, at which the radiative 
contribution from a pseudophotosphere in the optical rivals that 
from the nebula. 

\section{Conclusions}

The results of this {\em Letter} can be summarized as follows: 

(i)
Our photometry of Z\,And during its recent active phase 
(2000 -- 2003) revealed a~deep~minimum~in~the~$U,~B$~and~$V$~LCs. 

(ii)
The position of the minimum at the inferior conjunction of the 
cool component in the binary and the behaviour of colour indices 
strongly suggest that the minimum was caused by the eclipse of 
the active hot object by the red giant. 

(iii)
Physically plausible fit of the eclipse profile suggests 
a disk-like structure for the active hot object. Our analysis 
of the SED in the UV continuum during the previous (1984 -- 1986) 
active phase confirmed this result. The present knowledge of 
fundamental parameters of the system restricts the orbital 
inclination to 76$^{\circ}$ -- 90$^{\circ}$. 

(iv)
The presence of the Rayleigh attenuated far-UV continuum 
even during the quiescent phase at $\varphi \sim 0$ confirms 
the very high inclination of the Z\,And orbit. 

(v)
A dominance of the nebular component of radiation in the optical 
during the maximum and quiescence makes it difficult to observe 
the eclipse effect. It can be observed only at specific brightness 
phases. 

\begin{acknowledgements}
This work used in part the visual estimates of Z\,And gathered by 
the members of AFOEV available from the CDS database and CCD 
observations made by Makoto Watanabe and Dong West available from 
the VSNET database. 
This research has been supported by the Slovak Academy of Science 
under a grant No. 1157. 
\end{acknowledgements}


\begin{thebibliography}{}
\bibliographystyle{aa}

\bibitem[Cardelli et al. (1989)]{ccm89}
        Cardelli, J.A., Clayton, G.C., Mathis, J.S. 1989, ApJ, 345, 245 

\bibitem[FC95()]{f-c95}
         Fern\'andez-Castro, T., Gonz\'alez-Riestra, R., Cassatella, A., 
         Taylor,A.R., Seaquist E.R. 1995, ApJ, 442, 366 (FC95)

\bibitem[Formiggini \& Leibowitz (1994)]{fl94}
         Formiggini, L., Leibowitz, E.M. 1994, A\&A, 292, 534

\bibitem[Isliker et al. (1989)]{inv89}
         Isliker, H., Nussbaumer, H., Vogel, M. 1989, A\&A, 219, 271

\bibitem[Kva\v ckay (1994)]{kv94}
         Kva\v ckay, P. 1994, private communication

\bibitem[Mattei (1978)]{matt}
         Mattei, J.A. 1978, J. Roy. Astron. Soc. Can., 72, 61

\bibitem[M\"urset \& Schmid (1999)]{ms99}
         M\"urset, U., Schmid, H.M. 1999, A\&ASS, 137, 473

\bibitem[MK96()]{mk96}
         Mikolajewska, J., Kenyon, S.J. 1996, AJ, 112, 1659 (MK96)

\bibitem[NV89()]{nv89}
         Nussbaumer, H., Vogel, M. 1989, A\&A, 213, 137 (NV89)

\bibitem[Schmid \& Schild (1997)]{ss97}
         Schmid, H.M., Schild, H. 1997, A\&A, 327, 219

\bibitem[Skopal (1998)]{sk98}
	 Skopal, A. 1998, A\&A, 338, 599

\bibitem[Skopal (2001a)]{sk01a}
         Skopal, A. 2001a, A\&A, 366, 157

\bibitem[Skopal (2001b)]{sk01b}
         Skopal, A. 2001b, Contrib. Obs. Skalnate Pleso, 31, 119

\bibitem[Skopal (2003)]{sk03}
         Skopal, A. 2003, New Astron. (in press, astro-ph/0302542)

\bibitem[Skopal et al. (2000)]{sk_al00}
         Skopal, A., Chochol, D.,  Pribulla, T.,  Va\v{n}ko, M. 2000, 
         IBVS No. 5005

\bibitem[Skopal et al. (2002)]{sk_al02}
         Skopal, A., Va\v{n}ko, M., Pribulla, T., Wolf, M., Semkov, E., 
         Jones, A. 2002, Contrib. Obs. Skalnate Pleso, 32, 62

\bibitem[Sokoloski \& Bildsten (1999)]{sb99}
         Sokoloski, J.L., Bildsten, L. 1999, ApJ, 517, 919

\bibitem[Swings \& Struve (1941)]{ss41}
         Swings, P., Struve, O. 1941, ApJ, 93, 356

\bibitem[Tomov et al. (2003)]{ttt03}
         Tomov, N.A. Taranova, O., Tomova, M. 2003, A\&A (in press)

\end{thebibliography}
\end{document}